\documentclass[aps,prl,twocolumn,notitlepage]{revtex4-1}
\usepackage{epsfig,amssymb,amsmath,mathrsfs,amsthm,graphicx,float,color}

\usepackage{subfigure}

\newcommand{\R}{\mathbb{R}}

\newcommand{\set}[1]{\mathsf{#1}}
\newcommand{\grp}[1]{\mathsf{#1}}
\newcommand{\spc}[1]{\mathcal{#1}}

\def\d{{\rm d}}

\def\>{\rangle}
\def\<{\langle}

\newcommand{\map}[1]{\mathcal{#1}}
\newcommand{\Tr}{\operatorname{Tr}}

\newcommand{\St}{{\mathsf{St}}}

\newtheorem{theo}{Theorem}

\newtheorem{lemma}{Lemma}

\newtheorem{defi}{Definition}

\begin{document}

\title{Memory Effects in Quantum Metrology} 

\author{Yuxiang Yang} \affiliation{Institute for Theoretical Physics, ETH Z\"urich, 8093 Z\"urich, Switzerland}
\email{yangyu@phys.ethz.ch}
\begin{abstract}
Quantum metrology concerns estimating a parameter from multiple identical uses of a quantum channel.
We extend quantum metrology beyond this standard setting and consider estimation of a physical process with quantum memory, here referred to as a parametrized quantum comb.
We present a theoretic framework of metrology of quantum combs, and derive a general upper bound of the comb quantum Fisher information. 
The bound can be operationally interpreted as the quantum Fisher information of a memoryless quantum channel times a dimensional factor. 
We then show an example where the bound can be attained up to a factor of four.
With the example and the bound, we show that memory in quantum sensors plays an even more crucial role in the estimation of combs than in the standard setting of quantum metrology.
\end{abstract}

\maketitle

\noindent{\it Introduction.\ }
Steady developments in quantum communication and quantum memory allow us to measure a physical quantity with higher precision \cite{degen2017quantum,pirandola2018advances}.
By harnessing quantum control and ancillary memory qubits, adaptive metrological strategies can improve the performance of sensing even in the presence of noise \cite{dur2014improved,yuan2015optimal,sekatski2017quantum,zhou2018achieving}.

In standard quantum metrology, the goal is often to estimate identical copies of the same quantum gate, possibly subject to noise \cite{escher2011general,kolodynski2013efficient,chin2012quantum}, that are available either in parallel \cite{giovannetti2006quantum} or in arbitrary order \cite{van2007optimal,ji2008parameter}. 
With the advance of quantum technologies, however, the focus is transitioning 
to more complex and realistic settings, where the parameter to estimate is contained in a network or an adaptive physical process \cite{komar2014quantum,proctor2018multiparameter,ge2018distributed,eldredge2018optimal}.
In such settings, one would have to deal with a circuit with a complex underlying structure, or in a network equipped with communication channels and memories, which requires a model with a higher-order structure than parametrized quantum gates.

In this Letter, we extend quantum metrology beyond the standard setting of estimating parametrized quantum channels. We consider  estimating an unknown parameter from a physical process consisting in $K$ sequentially arranged parametrized quantum channels interconnected by quantum memory, here referred to as a (parametrized) $K$ comb.
We present a theoretic framework of quantum metrology with such parametrized combs, and derive a general upper bound of the $K$-comb quantum Fisher information (QFI). The QFI bound can be operationally interpreted as the quantum Fisher information of a memoryless channel times a dimensional factor that grows exponentially in $K$.
We then show an example of estimating a protected parameter, where the QFI bound can be attained up to a factor of four, revealing that the best possible precision of estimating a $K$ comb can decay exponentially with $K$.
 With the example and the QFI bound, we capture the power and the limitation of memory effects in quantum metrology, previously observed in the discrimination of quantum channels \cite{chiribella2008memory}.
In particular, memory in quantum sensors plays an even more crucial role in the estimation of quantum combs than in the standard setting of quantum metrology.

\medskip
\noindent{\it Quantum metrology in the presence of memory.\ }
In this Letter, our goal is to estimate an unknown parameter $\theta$ given access to a quantum machine that has its own memory, which belongs to a family of parametrized quantum machines $\{\map{N}_{\theta}\}_{\theta\in\set{\Theta}}$. 
This quantum machine could be, for instance, a quantum circuit with ancillary qubits or
a noisy physical process with an inaccessible environment.
To show memory effects in quantum metrology in the most straightforward way, we focus on single parameter estimation and $\set{\Theta}\subset\R$. It is convenient to characterize such a quantum machine $\map{N}_{\theta}$ by a parametrized quantum comb \cite{chiribella2008quantum,chiribella2009theoretical,bisio2011quantum}. 
A quantum comb is mathematically characterized by a positive operator, called the Choi operator \cite{choi1975completely}, that satisfies a series of normalization constraints. More details on quantum combs can be found in the original Letter \cite{chiribella2008quantum}.

\begin{figure}  [t!]
\begin{center}
  \includegraphics[width=0.9\linewidth]{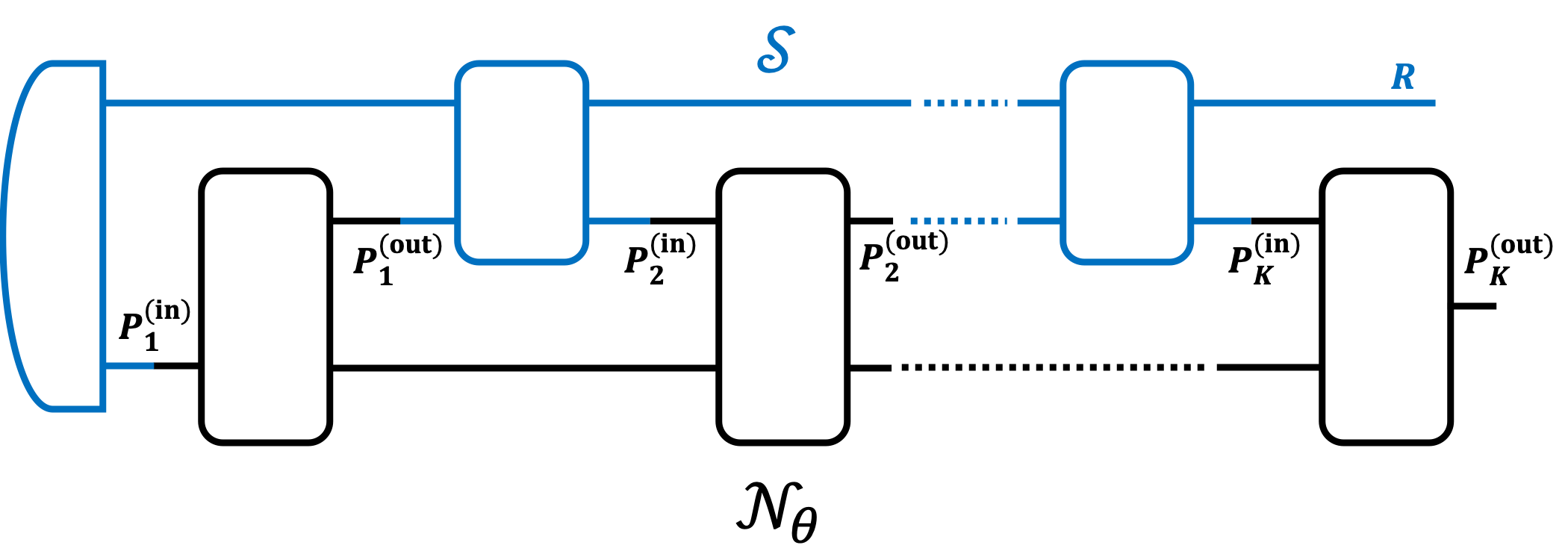}
  \end{center}
\caption{\label{fig-probe}
  {\bf Estimating a $K$ comb with a sensor.} In quantum comb metrology, the goal is to estimate a parameter $\theta$ from a $K$ comb $\map{N}_\theta$ consisting of $K$ phases (plotted in black). For this purpose, one uses a sensor $\map{S}$ (plotted in blue), which consists of preparing an input state and applying quantum gates in between the phases.}
\end{figure}

As illustrated in Fig.\ \ref{fig-probe}, the action of the comb $\map{N}_\theta$ is naturally divided into $K$ consecutive \emph{phases} $P_1,P_2,\dots,P_K$.
Each phase is connected to the next phase by a quantum memory (or quantum communication, if the phases are spatially separate).
 For each phase $P_n$, the comb takes a quantum state from an input port $P_n^{({\rm in})}$, performs a $\theta$-dependent quantum channel jointly on the input and the memory, and produces a quantum state from an output port $P_n^{({\rm out})}$. From now on, we will refer to it as a (parametrized) $K$ comb. 

In the standard context of quantum gate estimation, one prepares a quantum state and sends it through unknown gates. Here, to estimate a $K$ comb $\map{N}_\theta$, we need to connect it to a \emph{quantum sensor} $\map{S}$ with memory (hereafter referred to as a sensor), which could be a complex composition of quantum states, gates, memory, and communication channels.
Mathematically, the sensor $\map{S}$ is also modeled by a quantum comb, which takes $P_m^{({\rm out})}$ (for $m\in[K-1]$) as its input ports and $P_{n}^{({\rm in})}$ (for $n\in[K]$) as well as an ancillary system $R$ as its output ports. This means that a sensor is such a quantum comb that it ``eats" the $K$ comb $\map{N}_\theta$ and ``spits" a quantum state. 
Formally, this is taken care of by the link product \cite{chiribella2008quantum,chiribella2009theoretical,bisio2011quantum}, which is an operation to composite two combs  in a way that any input/output ports sharing the same tag are concatenated. In our case, the link product of the $K$ comb and the sensor results in a new quantum comb, denoted as $\map{N}_{\theta}\ast\map{S}$. As shown in Fig. \ref{fig-probe}, all ports $P_m^{({\rm out})}$ (for $m\in[K-1]$) and $P_{n}^{({\rm in})}$ (for $n\in[K]$) of the $K$ comb are connected to the corresponding ports of the sensor. Therefore, $\{\map{N}_{\theta}\ast\map{S}\}_{\theta\in\set{\Theta}}$ is in fact a family of quantum states on $\spc{H}_K^{({\rm out})}\otimes\spc{H}_{R}$. Here $\spc{H}_{n}^{({\rm out/in})}$ denotes the Hilbert space of $P_n^{({\rm out/in})}$.

A distinctive difference between quantum comb metrology and quantum gate estimation that makes the former more tricky to deal with is the \emph{memory effect} of both the sensor and the $K$ comb. Unlike quantum states that are used to probe parametrized gates, sensors are capable of memorizing the information on $\theta$ (possibly in the quantum form) at the end of each phase and send refined information back into the $K$ comb. Similarly, the $K$ comb is also capable of storing the input from the sensor in its underlying structure and making it interact with future inputs. Our goal is to see the impact of such effects on the best achievable precision.

Physically, the role of the sensor is to extract $\theta$ from the $K$ comb and to encode it into a quantum state. The QFI of quantum combs can be defined via the Fisher information of quantum states, by optimizing over all possible sensors:
\begin{defi}
The quantum Fisher information of $\{\map{N}_{\theta}\}$ is defined as 
\begin{align}
F_{\rm Q}\left[\map{N}_{\theta}\right]:=\max_{\map{S}}F_{\rm Q}\left[\map{N}_{\theta}\ast \map{S}\right],
\end{align}
where $F_{\rm Q}[\rho_\theta]$ denotes the QFI of a quantum state $\rho_\theta$ and the maximum is taken over all sensors $\map{S}$ such that $\map{N}_\theta\ast\map{S}$ is a quantum state. 
\end{defi}
With this definition of the comb QFI, we can apply the quantum Cram\'{e}r-Rao bound \cite{helstrom1976quantum,holevo2011probabilistic,braunstein1994statistical,hayashi2017quantum} and extend it to quantum combs: Denoting by $\delta\theta[\map{N}_\theta]$ the root-mean-square error of estimating $\theta$ from $\map{N}_\theta$, we have
\begin{align}
\delta\theta[\map{N}_\theta]&
=\min_{\map{S}}\left(\min_{\map{M}_{\hat{\theta}}}\delta\theta\left[\map{N}_{\theta}\ast\map{S}\ast\map{M}_{\hat{\theta}}\right]\right)\nonumber\\
&\ge\min_{\map{S}}\left(\frac{1}{\sqrt{\nu F_{\rm Q}\left[\map{N}_{\theta}\ast \map{S}\right]}}\right)\nonumber\\
&=\frac{1}{\sqrt{\nu F_{\rm Q}\left[\map{N}_{\theta}\right]}}
\end{align}
where in the first step the minimum is taken over all quantum estimators $\map{M}_{\hat{\theta}}$ that measure the state $\map{N}_{\theta}\ast\map{S}$ and output an unbiased estimate $\hat{\theta}$ of $\theta$, and $\nu$ is the number of repetitions of the experiment.

\medskip
\noindent{\it A general upper bound on the comb QFI.\ }
From the above discussion, we can see that the precision of estimating $\theta$ is determined by the QFI of a $K$ comb. However, the derivation of the QFI is not easy even for the simplest $K=1$ case, when the $K$ comb is reduced to a quantum channel.
A closed-form expression of the QFI was derived only for particular types of quantum channels (see, for instance, Refs.\ \cite{hayashi2011comparison,demkowicz2014using,takeoka2016optimal,pirandola2017ultimate}). Therefore, it is more sensible to look for an upper bound of the comb QFI so as to see the power and the limitation of metrology with a quantum comb.

In the following, we derive such an upper bound of the comb QFI, which implies a lower bound on the error of parameter estimation from the comb. We will use the abbreviation $\Psi:=|\Psi\>\<\Psi|$ of pure state density matrices and denote by $|\Phi^+\>$ a maximally entangled state $(1/\sqrt{d})\sum_j |j\>|j\>$.

\begin{figure}[t!]
\centering
\subfigure[]{\label{fig:postselect}
 \includegraphics[width=0.9\linewidth]{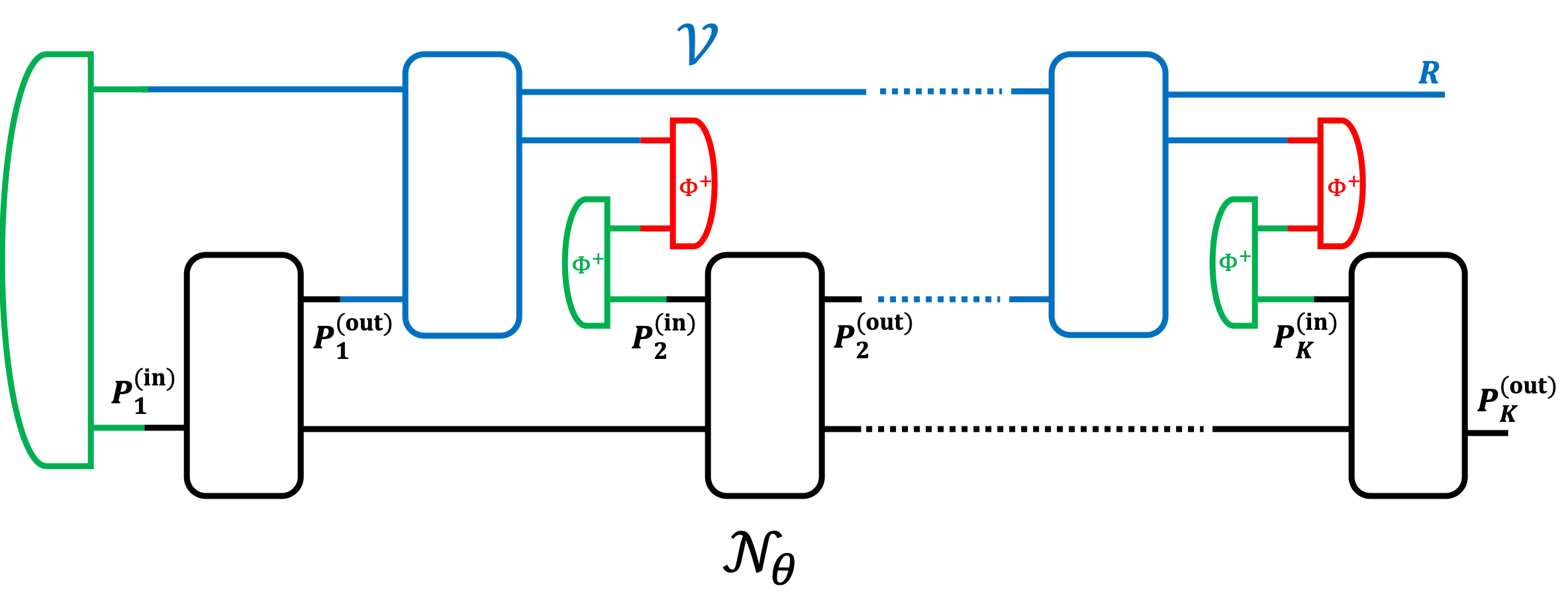}
}
\hfill
\subfigure[]{\label{fig:networkstate}
\includegraphics[height=3.5cm]{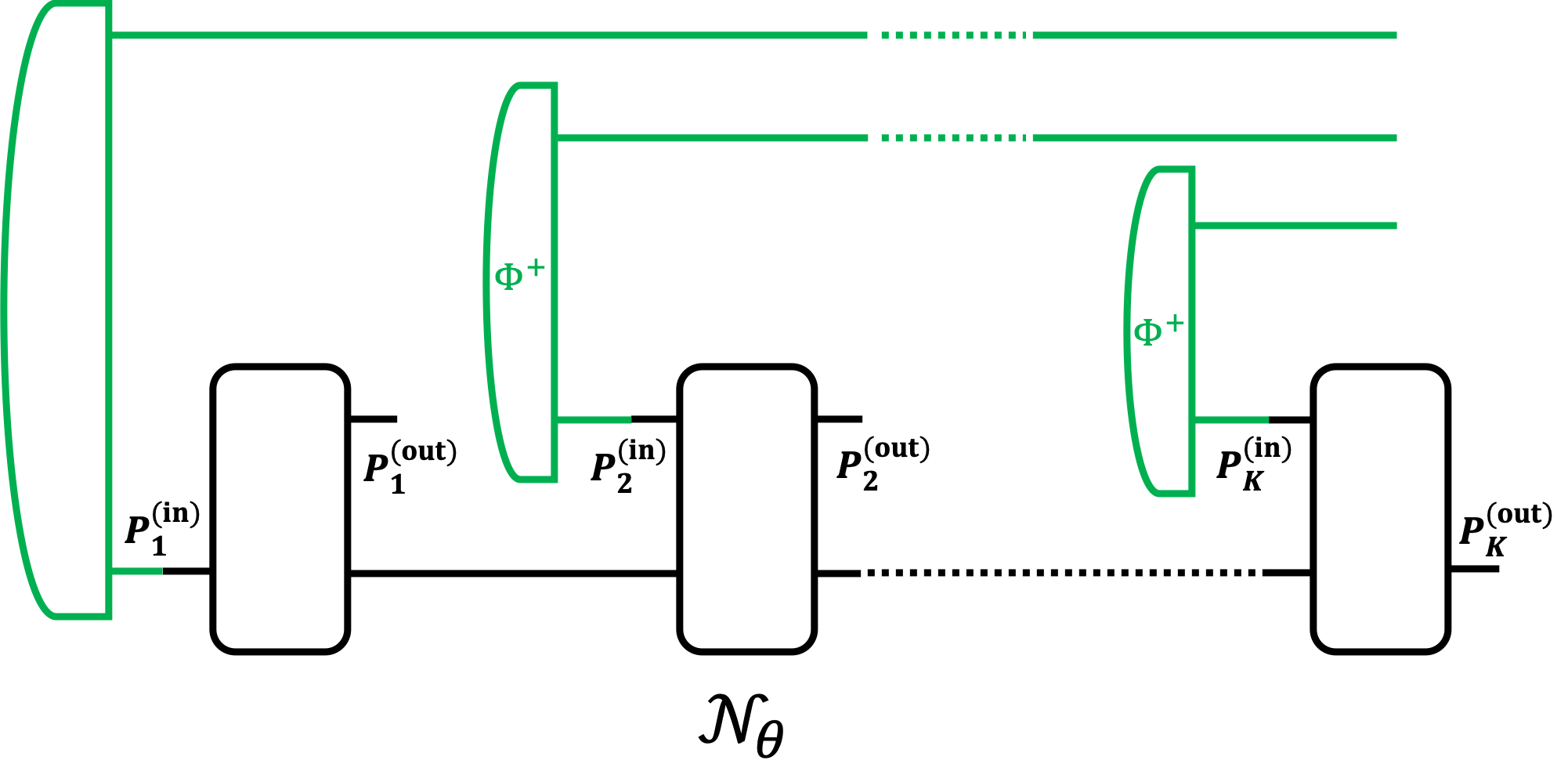}
}
\caption{{\bf Bounding the memory effect of $\map{N}_\theta$.} The wires connecting the sensor to the comb $\map{N}_\theta$ can be winded using probabilistic teleportation, as plotted in Fig.\ \ref{fig:postselect}. The QFI of $\map{N}_\theta$ can thus be bounded in terms of the QFI of the quantum state $\Phi_{\map{N}_{\theta}\ast\Psi_1}$, which is depicted in Fig.\ \ref{fig:networkstate}. }
\end{figure}

First, observe that any sensor can be decomposed as $\map{S}=\Psi_1\ast\map{V}$, where $\Psi_1\in\St(\spc{H}_1^{({\rm in})}\otimes\tilde{\spc{H}}_1^{({\rm in})})$ is a suitable input state (with $\tilde{\spc{H}}_1^{({\rm in})}\simeq \spc{H}_1^{({\rm in})}$ being a reference system)  and $\map{V}$ is a quantum comb.
An obstacle in determining $F_{\rm Q}\left[\map{N}_\theta\right]$ is that the information on $\theta$ can flow out of $\map{N}_\theta$ via an output port and then back into $\map{N}_\theta$ via subsequent input ports, owing to the memory effect of $\map{V}$.
To overcome this obstacle, we use a trick of postselection, which works by noticing that the action of a sensor $\map{S}$ is equivalent to the following probabilistic protocol, as depicted in Fig. \ref{fig:postselect}:
\begin{enumerate}
\item Send a proper state $\Psi_1$ into the first input port and a maximally entangled state $\Phi^+_n$ into the $n$-th input port for $n>1$, with $|\Phi^+_n\>$ being the maximally entangled state on $\spc{H}_n^{({\rm in})}\otimes\tilde{\spc{H}}_n^{({\rm in})}$ (with $\tilde{\spc{H}}_n^{({\rm in})}\simeq\spc{H}_n^{({\rm in})}$).
\item Feed all but the last of the output ports of $\map{N}_\theta\ast\Psi_1$ into a quantum comb $\map{V}$.
\item Perform a Bell test on the $n$-th output port of $\map{V}$ and the open end of $\Phi^+_n$ (for $n>1$); postselect the outcome $\Phi^+_n$. 
\end{enumerate}
Notice that we are abusing a bit the notation here, since the quantum comb $\map{V}$ has different (but isomorphic) output Hilbert spaces from the one in the decomposition of $\map{S}$. 

In view of estimating the parameter $\theta$, this probabilistic protocol can be regarded as one step of probabilistic metrology \cite{fiuravsek2006optimal,chiribella2013quantum,gendra2013quantum}, where one performs a measurement and postselects one particular outcome to enhance the performance of parameter estimation.
The limitation of probabilistic quantum metrology can be made clear via the following equation \cite[Eq.\ (30)]{combes2014quantum}:
For a family of parametrized quantum states $\{\rho_\theta\}$ and a quantum operation (i.\,e.\ a completely positive, trace-nonincreasing linear map) $\map{M}^{({\rm succ})}$, we have
\begin{align}\label{postselect}
F_{\rm Q}\left[\rho_\theta\right]\ge p^{({\rm succ})}_\theta\cdot F_{\rm Q}\left[\rho^{({\rm succ})}_{\theta}\right],
\end{align}
where $p^{({\rm succ})}_\theta:=\Tr\map{M}^{({\rm succ})}(\rho_\theta)$ is the success probability of the postselection and $\rho^{({\rm succ})}_{\theta}:=\map{M}^{({\rm succ})}(\rho_\theta)/p^{({\rm succ})}_\theta$ is the output state of $\map{M}^{({\rm succ})}$.

In our case, the probability that the teleportation of a $d$-dimensional system succeeds without a unitary correction is $d^{-2}$. The success probability of the postselection is thus
\begin{align}\label{succ-prob}
p^{({\rm succ})}_\theta=\left(\prod_{i=2}^K d^{({\rm in})}_i\right)^{-2},
\end{align}
where $d^{({\rm in})}_i$ is the dimension of the $i$-th output port. The comb QFI is thus bounded as
\begin{align}\label{bound-temp}
F_{\rm Q}\left[\Phi_{\map{N}_{\theta}\ast\Psi_1}\ast\map{V}\right]\ge p^{({\rm succ})}_\theta\cdot F_{\rm Q}\left[\map{N}_{\theta}\ast\map{S}\right],
\end{align}
where $\Phi_{\map{N}_{\theta}\ast\Psi_1}$ [see also Fig.\ \ref{fig:networkstate}] is the Choi state \cite{choi1975completely,chiribella2008quantum} of $\map{N}_{\theta}\ast\Psi_1$, defined as
\begin{align}\label{networkstate}
\Phi_{\map{N}_{\theta}\ast\Psi_1}:=\map{N}_{\theta}\ast\left(\Psi_1\otimes\Phi^+_2\otimes\cdots\otimes\Phi^+_{K}\right).
\end{align}
Moreover, the output ports of $\map{V}$ are now detached from the input ports of $\map{N}_\theta$ thanks to the postselection. 
Noticing that data processing does not increase distinguishability \cite[Chapter 6]{hayashi2017quantum}, to maximize the comb QFI we should always take the quantum comb $\map{V}$ to be a sequence of isometries. Under this condition, we have
\begin{align}\label{eq-QFI}
F_{\rm Q}\left[\Phi_{\map{N}_{\theta}\ast\Psi_1}\ast\map{V}\right]= F_{\rm Q}\left[\Phi_{\map{N}_{\theta}\ast\Psi_1}\right].
\end{align}

Combining Eqs.\ (\ref{succ-prob}), (\ref{bound-temp}), and (\ref{eq-QFI}), we get that the QFI of $\map{N}_{\theta}\ast\map{S}$ is bounded by $\left(\prod_{i=2}^{K}d^{({\rm in})}_{i}\right)^2$ times the QFI of the quantum state $\Phi_{\map{N}_{\theta}\ast\Psi_1}$ [see Fig.\ \ref{fig:networkstate}].
In summary, we derived the following theorem:
\begin{theo}\label{thm-bound}
The QFI of a $K$ comb $\map{N}_\theta$ is upper bounded as
\begin{align}\label{bound}
F_{\rm Q}\left[\map{N}_{\theta}\right]\le \left(\prod_{i=2}^{K}d^{({\rm in})}_{i}\right)^2 \max_{\Psi_1}F_{\rm Q}\left[\Phi_{\map{N}_{\theta}\ast\Psi_1}\right],
\end{align}
where $\Phi_{\map{N}_{\theta}\ast\Psi_1}$ is defined in Eq.\ (\ref{networkstate}) and the maximum is taken over all $\Psi_1\in\St(\spc{H}_1^{({\rm in})}\otimes\tilde{\spc{H}}_1^{({\rm in})})$ with $\tilde{\spc{H}}_1^{({\rm in})}\simeq \spc{H}_1^{({\rm in})}$ being a reference system.
\end{theo}

Theorem \ref{thm-bound} provides an upper bound on the QFI of an arbitrary quantum comb. The only optimization required in the upper bound (\ref{bound}) is the maximization of the QFI over input states to the first port. 
Notice that, since the quantum comb $\map{N}_{\theta}\ast\left(\Phi^+_2\otimes\cdots\otimes\Phi^+_{K}\right)$ has only one input port, the optimization is equivalent to finding the QFI of a quantum channel.

The QFI term on the right hand side of Eq.\ (\ref{bound}) is attained by a \emph{phase-parallel} scheme, namely by feeding a quantum state into each of the input ports and collecting the states from the output ports for measurement. 
In a phase-parallel scheme, the comb $\map{N}_\theta$ is treated as a quantum channel with a multipartite input.
The information on $\theta$ never flows back into the comb $\map{N}_\theta$.
Therefore, the comb QFI bound (\ref{bound}) shows that the memory effect of a sensor improves the sensitivity of parameter estimation at most by a factor exponential in $K$.
Note that a phase-parallel scheme is not necessarily local, since a phase of $\map{N}_\theta$ may consist of joint operations on multiple physical nodes in a realistic quantum network.

It is the main objective of quantum metrology to compare different strategies in terms of the asymptotic scalings of their QFIs with respect to the amount of required resources.  
In the conventional setting of parallel gate estimation \cite{giovannetti2004quantum,giovannetti2006quantum}, for instance, the resource is taken to be the number of calls to the parametrized quantum gate. Here in a quantum comb, the resource is quantified by the number of phases $K$.
An obvious observation is that the scaling suggested by the bound (\ref{bound}) is distinct from what one encounters when estimating a $K$ comb that does not have a memory. 
For instance, if one gets back to the standard context of quantum metrology and considers each phase to be an individual quantum channel $\map{C}_\theta$, the scaling of the comb QFI is at most the Heisenberg scaling $K^2$ if one applies a suitable adaptive strategy \cite{de2005quantum,higgins2007entanglement}, whereas a phase-parallel scheme achieves the standard quantum limit scaling $K$.
In Eq.\ (\ref{bound}), in contrast, if all the input dimensions are equal to $d$, we get
\begin{align}
F_{\rm Q}\left[\map{N}_{\theta}\right]\le d^{2(K-1)}\cdot F_{\rm Q}\left[\map{N}_{\theta},{\rm phase-parallel}\right],
\end{align}
where $F_{\rm Q}\left[\map{N}_{\theta},{\rm phase-parallel}\right]$ corresponds to the optimization term in Eq.\ (\ref{bound}).
Next, we present a scenario of quantum metrology in the presence of memory effects, where the above bound is saturated up to a constant factor.

\begin{figure}  [t]
\begin{center}
  \includegraphics[width=\linewidth]{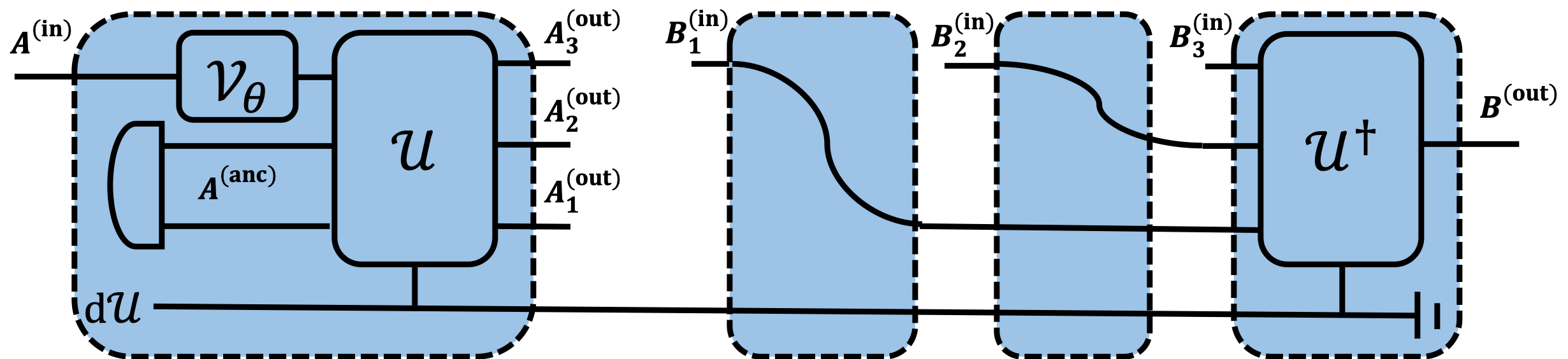}
  \end{center}
\caption{\label{fig-unitary}
  {\bf Estimating a protected parameter.} An experimenter is to estimate a protected parameter $\theta$ from a $K$ comb, consisting of $K$ blue boxes (whose latent structure is invisible) with input and output ports. The first box performs a parametrized gate $\map{V}_\theta$, and mixes the resultant state with a fixed state by a random unitary $\map{U}$ that serves as the shield. Accessing the remaining $K-1$ boxes jointly will result in a key unitary $\map{U}^\dag$, which recovers the information on $\theta$. If the experimenter fails to do so, the information on $\theta$ will be deteriorated.}
\end{figure}

\medskip
\noindent{\it Estimating a protected parameter.\ }
We now consider a scenario where the optimal strategy is exponentially more efficient than the phase-parallel strategy in Eq.\ (\ref{bound}). 
As illustrated in Fig.\ \ref{fig-unitary}, the $K$ comb $\map{N}_\theta$ in this scenario encodes a parameter $\theta$ in a unitary gate, and then uses a shield-key system to protect the parameter. We label the first phase as $A$ and the remaining phases as $\{B_i\}_{i\in[K-1]}$, and we shall use the convention $\map{U}(\cdot):=U(\cdot)U^\dag$ for a unitary.
In the first phase, the input state goes through a parametrized unitary $\map{V}_\theta$ and then it is mixed up with an ancillary state by a shielding unitary $\map{U}$ sampled with the Haar measure $\d U$ of $\grp{SU}(D)$. In the $n$-th phase ($2\le n\le K-1$), the comb simply stores the input state and outputs nothing.
In the $K$-th (last) phase, the comb merges the input with all the previously stored states, and performs the key unitary $\map{U}^\dag$ jointly on all these input states.

Here we consider the case when $D=\prod_{i=1}^{K-1} d_{A_i^{({\rm out})}}$ is an even number, and $d_{A^{({\rm in})}}=2$.  For the dimensions to match, we have $d_{A_n^{({\rm out})}}=d_{B_n^{({\rm in})}}$ for $n\in[K-1]$ and
\begin{align}\label{dimension}
D=\prod_{i=1}^{K-1} d_{B_i^{({\rm in})}}.
\end{align}
The ancillary system, denoted as $A^{({\rm anc})}$, has dimension $D/2$, and we assume its state to be the maximally mixed state.


Denote by $\Psi$ a quantum state in the Hilbert space of $A^{({\rm in})}$ and a reference system $\rm R$ such that the QFI of $\Psi_\theta:=\left(\map{V}_\theta\otimes\map{I}_{\rm R}\right)(\Psi)$, with $\map{I}_{\rm R}$ denoting the identity channel on the reference, achieves the maximum. 
The optimal scheme is to send (perhaps a part of) $\Psi$ through the input port $A^{({\rm in})}$ and then to connect $A_n^{({\rm out})}$ to the corresponding input port $B_n^{({\rm in})}$ for every $n\in[K-1]$. With this approach, the shield $\map{U}$ cancels out with the key $\map{U}^\dag$, and thus the optimal QFI for this comb is
\begin{align}\label{QFI-optimal}
F_{\rm Q}\left[\map{N}_{\theta}\right]=F_{\rm Q}\left[\Psi_\theta\right].
\end{align}

Obviously, the optimal scheme requires a quantum memory in between different phases. We now compare it to the phase-parallel scheme corresponding to the right hand side term of Eq.\ (\ref{bound}). For this purpose, we need to evaluate $F_{\rm Q}\left[\Phi_{\map{N}_{\theta}\ast\Psi'}\right]$, which is the QFI of the output state when the comb is fed with the state $\Psi'_{A^{({\rm in})}\tilde{A}^{({\rm in})}}\otimes\Phi^+_{B^{({\rm in})}_1\tilde{B}^{({\rm in})}_1}\otimes\cdots\otimes\Phi^+_{B^{({\rm in})}_{K-1}\tilde{B}^{({\rm in})}_{K-1}}$. 
For conciseness, we denote by $B$ the $K-1$ phases $B_1,\dots,B_{K-1}$, and the input state can be rewritten as $\Psi'_{A^{({\rm in})}\tilde{A}^{({\rm in})}}\otimes\Phi^+_{B^{({\rm in})}\tilde{B}^{({\rm in})}}$.  
The output state for this input can be derived by invoking symmetry properties of the comb. Because of the twirling effect of the comb, $\tilde{A}^{({\rm in})}$ will be partially entangled with a two-dimensional subspace $B^{({\rm in},1)}$ of $B^{({\rm in})}$.
Explicitly, the output state reads
\begin{align}
&\Phi_{\map{N}_{\theta}\ast\Psi}=\left(\frac{\Psi_\theta}{D^2}\right)_{B^{({\rm in},1)}\tilde{A}^{({\rm in})}}\otimes\Phi^+_{A^{({\rm tot})}\tilde{B}^{({\rm in})}}\otimes\pi_{B^{({\rm in},2)}}\\
&+\left(\frac{(D^2-4)\rho_{\Psi'}+3\pi^-_{\theta,\Psi'}}{D^2}\right)_{\tilde{A}^{({\rm in})}B^{({\rm in},1)}}\otimes\pi^-_{A^{({\rm tot})}\tilde{B}^{({\rm in})}}\otimes\pi_{B^{({\rm in},2)}}\nonumber
\end{align}
where $\pi$ is the maximally mixed state, $\pi^-:=(I-\Phi^+)/(D^2-1)$ for a bipartite system, $(\rho_{\Psi'})_{AB}:=(I_A\otimes\Tr_A\Psi')/d_A$ is a $\theta$-independent state, $\pi^-_{\theta,\Psi'}:=(4\rho_{\Psi'}-\Psi'_\theta)/3$ is a $\theta$-dependent state, $A^{({\rm tot})}$ is the joint system of $A^{({\rm in})}$ and $A^{({\rm anc})}$, and $B^{({\rm in},2)}$ is defined via the decomposition $B^{({\rm in})}\simeq B^{({\rm in},1)}\otimes B^{({\rm in},2)}$. The derivation is left to the Appendix.

Next, we apply the convexity of the QFI of states, i.\,e., $F_{\rm Q}[p\rho_\theta+(1-p)\sigma_\theta]\le pF_{\rm Q}[\rho_\theta]+(1-p)F_{\rm Q}[\sigma_\theta]$ for two parametrized state families $\{\rho_\theta\}$ and $\{\sigma_\theta\}$  (see, for instance, \cite[Chapter 6]{hayashi2017quantum}).
Using the convexity and observing that $\rho_{\Psi'}$ has no QFI, we can bound the QFI of the output state as
\begin{align}
F_{\rm Q}\left[\Phi_{\map{N}_{\theta}\ast\Psi}\right]&\le\frac{1}{D^2}F_{\rm Q}[\Psi'_\theta]+F_{\rm Q}\left[\frac{(D^2-4)\rho_{\Psi'}+3\pi^-_{\theta,\Psi'}}{D^2}\right] \nonumber\\
&\le\frac{1}{D^2}F_{\rm Q}[\Psi'_\theta]+\frac{3}{D^2}F_{\rm Q}\left[\pi^-_{\theta,\Psi'}\right].
\label{output-QFI1}
\end{align}
We remark that the first inequality is actually an equality since $\pi^-$ and $\Phi^+$ are orthogonal to each other.

Since $\Psi$ is such an input state that the QFI $F_{\rm Q}[\Psi_\theta]$ attains its maximum,
from Eq.\ (\ref{output-QFI1}) we get
\begin{align}\label{QFI-parallel}
F_{\rm Q}\left[\Phi_{\map{N}_{\theta}\ast\Psi'}\right]\le\frac{4}{D^2}F_{\rm Q}\left[\Psi_\theta\right].
\end{align}
Finally, we get the relation between the QFI of $\map{N}_\theta$ and the QFI of $\Phi_{\map{N}_{\theta}\ast\Psi}$ as
\begin{align}
F_{\rm Q}\left[\map{N}_{\theta}\right]\ge\frac14\left(\prod_{i=1}^{K-1} d_{B_i^{({\rm in})}}\right)^2F_{\rm Q}\left[\Phi_{\map{N}_{\theta}\ast\Psi}\right],
\end{align}
by substituting Eq.\ (\ref{QFI-parallel}) and Eq.\ (\ref{dimension}) into Eq.\ (\ref{QFI-optimal}). This clearly shows that the bound (\ref{bound}) is tight up to a factor of four for this scenario of estimating a protected phase.
Memory effects in quantum metrology are thus manifested by the fact that the optimal adaptive sensor is exponentially more powerful than the phase-parallel sensor.

\medskip
\noindent{\it Conclusion.\ } 
We established a framework for quantum comb metrology and showed the effect of memory in quantum metrology.
This could be the start of a new research direction, which deals with metrology in a fully quantum network or estimation of non-Markovian processes \cite{pollock2018non}. Our work also fits the current trend of studying quantum information processing of higher-order structures \cite{gour2018entropy,yuan2018hypothesis,liu2019resource}. 
We conclude with a remark that the quantum combs considered here have definite causal structures, while it was recently shown that interesting phenomena can be observed in quantum communication networks with an indefinite causal structure \cite{ebler2018enhanced}. It is thus meaningful to ask whether these phenomena extend to metrology, which may be related to probing an unknown spacetime structure.

\medskip
\begin{acknowledgements}
The author is grateful to Giulio Chiribella, Joseph M. Renes, and anonymous referees for valuable comments, to Masahito Hayashi for discussions, and to Mark Wilde for suggesting references. This work is supported by the Swiss National Science Foundation via the National Center for Competence in Research ``QSIT" as well as via project No.\ 200020\_165843 and by the Hong Kong Research Grant Council through Grant No.\ 17300918. 
\end{acknowledgements}

\bibliography{ref}
\bibliographystyle{unsrt}

\appendix

\begin{widetext}

\section{Derivation of Eq.\ (12).}

We can write any input state to the first port as $|\Psi'\>_{A^{({\rm in})}\tilde{A}^{({\rm in})}}=(I_{A^{({\rm in})}}\otimes X)|\Phi^+\>_{A^{({\rm in})}\tilde{A}^{({\rm in})}}$, where $X$ is a matrix constrained by $\Tr X^\dag X=d_{A^{({\rm in})}}=2$ and $I_{A^{({\rm in})}}$ is the identity matrix on $A^{({\rm in})}$. Since the ancillary system is in the maximally mixed state $\pi_{A^{({\rm anc})}}:=I_{A^{({\rm anc})}}/d_{A^{({\rm anc})}}$ with $d_{A^{({\rm anc})}}=D/2$, we first consider its purification $\Phi^+_{A^{({\rm anc})}\tilde{A}^{({\rm anc})}}$ and trace out $\tilde{A}^{({\rm anc})}$ in the end.
The output state is thus 
\begin{align*}
\Phi_{\map{N}_{\theta}\ast\Psi}=&\int\d U\,\left(\map{U}_{A^{({\rm in})}A^{({\rm anc})}}\map{V}_{\theta,A^{({\rm in})}}\otimes\map{X}_{\tilde{A}^{({\rm in})}}\otimes\Tr_{\tilde{A}^{({\rm anc})}}\otimes\ \map{U}^\dag_{B^{({\rm in})}}\otimes\map{I}_{\tilde{B}^{({\rm in})}}\right)\left(\Phi^+_{(A^{({\rm in})}A^{({\rm anc})})(\tilde{A}^{({\rm in})}\tilde{A}^{({\rm anc})})}\otimes\Phi^+_{B^{({\rm in})}\tilde{B}^{({\rm in})}}\right),
\end{align*}
where $\Phi^+_{(A^{({\rm in})}A^{({\rm anc})})(\tilde{A}^{({\rm in})}\tilde{A}^{({\rm anc})})}$ denotes the maximally entangled state between the product systems $A^{({\rm in})}A^{({\rm anc})}$ and $\tilde{A}^{({\rm in})}\tilde{A}^{({\rm anc})}$ and $\Tr_{\tilde{A}^{({\rm anc})}}$ denotes the partial trace operation.
Using the elementary property 
\begin{align}\label{trivial}
(Y\otimes I)|\Phi^+\>=(I\otimes Y^T)|\Phi^+\>
\end{align}
that holds for any matrix $Y$, the output state can be rewritten as
\begin{align}
\Phi_{\map{N}_{\theta}\ast\Psi}&=\int\d U\,\left(\map{U}_{A^{({\rm in})}A^{({\rm anc})}}\map{V}_{\theta,A^{({\rm in})}}\otimes\map{X}_{\tilde{A}^{({\rm in})}}\otimes\Tr_{\tilde{A}^{({\rm anc})}}\otimes\ \map{I}_{B^{({\rm in})}}\otimes\map{\bar{U}}_{\tilde{B}^{({\rm in})}}\right)\left(\Phi^+_{(A^{({\rm in})}A^{({\rm anc})})(\tilde{A}^{({\rm in})}\tilde{A}^{({\rm anc})})}\otimes\Phi^+_{B^{({\rm in})}\tilde{B}^{({\rm in})}}\right)\nonumber\\
&=\left(\map{T}_{(A^{({\rm in})}A^{({\rm anc})})\tilde{B}^{({\rm in})}}\otimes\map{X}_{\tilde{A}^{({\rm in})}}\map{V}_{\theta,\tilde{A}^{({\rm in})}}^T\otimes\Tr_{\tilde{A}^{({\rm anc})}}\otimes\ \map{I}_{B^{({\rm in})}}\right)\left(\Phi^+_{(A^{({\rm in})}A^{({\rm anc})})(\tilde{A}^{({\rm in})}\tilde{A}^{({\rm anc})})}\otimes\Phi^+_{B^{({\rm in})}\tilde{B}^{({\rm in})}}\right),\label{output-inter}
\end{align}
where $\map{T}$ is a twirling channel $\map{T}(\rho):=\int\d U\,\left(U\otimes \bar{U}\right)\rho\left(\bar{U}\otimes U\right)$ and $\bar{U}$ denotes the conjugate of $U$.

By Schur's lemma \cite{fulton2013representation}, the output of the twirling channel is of the form $\map{T}(\rho)=\Tr[\Phi^+ \rho]\Phi^++\Tr[(I-\Phi^+)\rho]\pi^-$, 
where $\pi^-:=\frac{1}{D^2-1}(I-\Phi^+)$ for a $(D\otimes D)$ bipartite system. 
Then we obtain the action of the twirling $\map{T}_{(A^{({\rm in})}A^{({\rm anc})})\tilde{B}^{({\rm in})}}$ on $\Phi^+_{(A^{({\rm in})}A^{({\rm anc})})(\tilde{A}^{({\rm in})}\tilde{A}^{({\rm anc})})}\otimes\Phi^+_{B^{({\rm in})}\tilde{B}^{({\rm in})}}$ as
\begin{align}
&\left(\map{T}_{(A^{({\rm in})}A^{({\rm anc})})\tilde{B}^{({\rm in})}}\otimes\map{I}_{\tilde{A}^{({\rm in})}\tilde{A}^{({\rm anc})}B^{({\rm in})}}\right)\left(\Phi^+_{(A^{({\rm in})}A^{({\rm anc})})(\tilde{A}^{({\rm in})}\tilde{A}^{({\rm anc})})}\otimes\Phi^+_{B^{({\rm in})}\tilde{B}^{({\rm in})}}\right)\nonumber\\
=&\frac{1}{D^2}\left(\Phi^+_{(A^{({\rm in})}A^{({\rm anc})})\tilde{B}^{({\rm in})}}\otimes\Phi^+_{(\tilde{A}^{({\rm in})}\tilde{A}^{({\rm anc})})B^{({\rm in})}}\right)+\left(\frac{D^2-1}{D^2}\right)\left(\pi^-_{(A^{({\rm in})}A^{({\rm anc})})\tilde{B}^{({\rm in})}}\otimes\pi^{-}_{(\tilde{A}^{({\rm in})}\tilde{A}^{({\rm anc})})B^{({\rm in})}}\right).\label{twirling}
\end{align}
Operationally, this twirling $\map{T}_{(A^{({\rm in})}A^{({\rm anc})})\tilde{B}^{({\rm in})}}$  swaps the entanglement with $A^{({\rm in})}A^{({\rm anc})}$ from $\tilde{A}^{({\rm in})}\tilde{A}^{({\rm anc})}$ to $\tilde{B}^{({\rm in})}$ with a success probability $1/D^2$, by performing correlated random unitaries. 

Now, we consider the action of the partial trace $\Tr_{\tilde{A}^{({\rm anc})}}$ operation on the state in Eq.\ (\ref{twirling}).
Note that $B^{({\rm in})}$ is isomorphic to $B^{({\rm in},1)}\otimes B^{({\rm in,2})}$ with $B^{({\rm in},1)}\simeq\tilde{A}^{({\rm in})}$ and $B^{({\rm in},2)}\simeq\tilde{A}^{({\rm anc})}$, $\Tr_{\tilde{A}^{({\rm anc})}}\Phi^+_{(\tilde{A}^{({\rm in})}\tilde{A}^{({\rm anc})})B^{({\rm in})}}=\Phi^+_{\tilde{A}^{({\rm in})}B^{({\rm in},1)}}\otimes\pi_{B^{({\rm in},2)}}$, and $\Tr_{\tilde{A}^{({\rm anc})}}\pi^-_{(\tilde{A}^{({\rm in})}\tilde{A}^{({\rm anc})})B^{({\rm in})}}=(D^2-1)^{-1}((D^2/4)I-\Phi^+)_{\tilde{A}^{({\rm in})}B^{({\rm in},1)}}\otimes\pi_{B^{({\rm in},2)}}$.
With this and substituting Eq.\ (\ref{twirling}) into Eq.\ (\ref{output-inter}), we obtain the output state as
\begin{align}
\Phi_{\map{N}_{\theta}\ast\Psi}&=\frac{1}{D^2}\Phi^+_{(A^{({\rm in})}A^{({\rm anc})})\tilde{B}^{({\rm in})}}\otimes\left(\map{X}_{\tilde{A}^{({\rm in})}}\map{V}_{\theta,\tilde{A}^{({\rm in})}}^T\otimes\map{I}_{B^{({\rm in},1)}}\right)\left(\Phi^+_{\tilde{A}^{({\rm in})}B^{({\rm in,1})}}\right)\otimes\pi_{B^{({\rm in},2)}}\nonumber\\
&\qquad+\frac{1}{D^2}\pi^-_{(A^{({\rm in})}A^{({\rm anc})})\tilde{B}^{({\rm in})}}\otimes\left(\map{X}_{\tilde{A}^{({\rm in})}}\map{V}_{\theta,\tilde{A}^{({\rm in})}}^T\otimes\map{I}_{B^{({\rm in},1)}}\right)\left(\frac{D^2}{4}I_{\tilde{A}^{({\rm in})}B^{({\rm in},1)}}-\Phi^+_{\tilde{A}^{({\rm in})}B^{({\rm in},1)}}\right)\otimes\pi_{B^{({\rm in},2)}}\nonumber\\
&=\frac{1}{D^2}(\Psi_\theta)_{B^{({\rm in},1)}\tilde{A}^{({\rm in})}}\otimes\Phi^+_{(A^{({\rm in})}A^{({\rm anc})})\tilde{B}^{({\rm in})}}\otimes\pi_{B^{({\rm in},2)}}\nonumber\\
&\qquad+\frac{1}{D^2}\left(\frac{D^2-4}{4}(I\otimes XX^\dag)+3\pi^-_{\theta,\Psi'}\right)_{\tilde{A}^{({\rm in})}B^{({\rm in},1)}}\otimes\pi^-_{(A^{({\rm in})}A^{({\rm anc})})\tilde{B}^{({\rm in})}}\otimes\pi_{B^{({\rm in},2)}}\nonumber
\end{align}
having used Eq.\ (\ref{trivial}), $d_{A^{({\rm in})}}=2$, and the relation $|\Psi'\>_{B^{({\rm in})}\tilde{A}^{({\rm in})}}=(I_{B^{({\rm in})}}\otimes X)|\Phi^+\>_{B^{({\rm in})}\tilde{A}^{({\rm in})}}$ in the last step. Here $\pi^-_{\theta,\Psi'}:=(I\otimes X)(I\otimes I-\Phi^+_\theta)(I\otimes X^\dag)/3$ is a quantum state.
Finally, using $XX^\dag/2=\Tr_{B^{({\rm in,1})}}\Psi'$ we get Eq.\ (12).

\end{widetext}

\end{document}